# A Graphical Tool for Testing Timed Systems based on Meta-Modeling and Graph Grammars

Hiba Hachichi[1], Ilham Kitouni[1], Kenza Bouaroudj[1] and Djamel-Eddine Saidouni[1]

[1] MISC Laboratory, University Mentouri
Constantine, 25000, Algeria

**Abstract**

The test is one of the approaches commonly used for validating systems to ensure qualitative and quantitative implementation requirements. In this paper, we interest in formal testing using graph transformation, thus we propose an approach for translating a Durational Actions Timed Automata model (DATA*) with a high number of states into a timed refusals region graph (TRRG) for creating a canonical tester and generating test cases using graph transformation. Though, our approach allows to generate automatically a visual modeling tool for DATA*, TRRG and the canonical tester. The cost of building a visual modeling tool from scratch is prohibitive. Meta-modeling approach is useful to deal with this problem since it allows the modeling of the formalisms themselves, by means of graph grammars. The meta-modeling tool AToM$^3$ is used.

## 1. Introduction

In recent years, technological progress in the fields of computer networks, telecommunications and multimedia systems can be considered as revolution which has a direct impact on our daily lives. This kind of systems is known by their complexity.

Formal testing can greatly increase the confidence in the functioning of these systems. It allows checking the correctness of a system with respect to its specification.

In this work we are interested in formal testing approach [17], [10] where the temporal behavior of systems is taken into account. This approach is based on timed refusals. Testing based on timed refusals allows the comparison between the behavior of the specification and the implementation, if the implementation refuses an action after each timed trace, the specification also refuses this action. That means *I* and *S* have the same timed traces and the same refusals sets. This theoretical approach is necessary to generate a canonical tester.

In this paper we use timed refusals region graph structure (TRRG). This structure allowed us to generate a canonical tester and test cases by making several transformations on it.

TRRG is constructed by applying transformation on Durational Actions Timed Automata model (DATA*) with a high number of states.

DATA* is a timed model. Its semantics expresses the durations of actions and other notions for specifying the real-time systems such as urgency and deadlines [4].This model is based on maximality semantics [18] and advocates the true concurrency; from this point of view, it is well suitable for modeling real time, concurrent and distributed systems.

In this paper, we propose firstly a program written in python language that transforms a DATA* structure, presented as a dotty file, to a DATA* structure written in the form of a python file respecting the syntax of AToM$^3$. The aim of this transformation is to consider DATA* structures with a high number of states. Secondly, we propose an approach and a tool for transforming DATA* into timed refusals region graph (TRRG). After, we transform this TRRG into canonical tester and we generate test cases using graph transformation [3], [13]. Indeed, we propose a DATA* meta-model and a TRRG /canonical tester meta-model. We use the meta-Modeling tool AToM$^3$ [2], [5] to generate automatically a visual Modeling tool to process models in DATA*, TRRG and the canonical tester. We also define a graph grammar to translate models presented above.

This paper is organized as follows: section 2 outlines some related work. In section 3 we recall some basic concepts about DATA*, TRRG, canonical tester and graph transformation. In section 4 we describe our approach. In section 5, we illustrate our approach through an example. The final section concludes the paper and gives some perspectives.

## 2. Related Work

This paper deals with formal testing approach and model graph transformations.

Firstly, we present several proposed works to tackle the problem of testing timed systems. Each of these works faces the problem from a different point of view. For instance, [16] uses the Extended Time Input Output State

Machine to develop an algorithm for creating a canonical tester It is used after to generate timed test sequences. In [17] they propose a fully automatic method for generating a real-time test sequences from a restricted sub class of timed automata called event-recording automata which restricts how clocks are reset in dense time context. This approach is based on de Nicola and Hennessy testing theory. A selection technique of timed tests is presented. This technique is based on symbolic analysis and coverage of a coarse equivalence class partitioning of the state space. The proposed conformance relation is a must/may preorder relation. In [6] authors present technique to test real-time systems through the derivation of executable test cases on a specification modeled as a timed automata, this study deals with an equivalent representation of timed automata: Clock region graphs [1]. A test purpose is modeled by an acyclic graph: All paths of this graph which are found on the specification will be considered as test cases. [15] presents a framework for black-box conformance testing of real-time systems. Specifications are modeled as timed automata and algorithms are proposed to generate two types of tests for this setting: Analog-clock tests, which measure dense time precisely, and digital-clock tests, which measure time with a periodic clock. A heuristic to generate test cases that covers all specification edges is briefly discussed.

Secondly we present some proposed tools in addition to AToM[3] [5] that used meta-Modeling concepts and visual tools like Generic Modeling environment (GME) [8], MetaEdit+ [12] and other tools from the Eclipse Generative Modeling tools (GMT) project such as Eclipse Modeling Framework (EMF), Graphical Modeling Framework (GMF) and Graphical Editing Framework (GEF) [7] . There are also similar tools which manipulate models by means of graph grammars and none of these has its own meta-Modeling layer, such as PROGRES, GReAT and AGG.

## 3. Background

In this paper, we use a testing structure named timed refusals region graph (TRRG). This graph allows us to generate a canonical tester and extract automatically test cases. TRRG, Canonical tester and test cases are obtained after applying several transformations on graph specification. In our case specifications are modeled by Durational Actions Timed Automata (DATA*) with a high number of states.

The transformation process is performed by a graph grammar that takes the DATA* model as an input, executes the rules of the grammar, and generates the canonical tester as output passing by TRRG.

In the following, we recall some basic notions about DATA* model, TRRG, canonical tester and graph transformations.

### 3.1 DATA* Model

The DATA* model (Durational Actions Timed Automata) [4] is a timed model defined by a timed transitions system over an alphabet representing actions to be executed. This model takes into account, in the specification, the duration of actions based on an intuitive idea: temporal and structural non-atomicity of actions. This model seems interesting and funneling more and more research because it coated models of timed automata by maximality semantics [18].

The DATA* model, as the temporized models takes in charge the notions of urgency and deadlines as temporal constraints of the system. Fig.1 illustrates an example of this model:

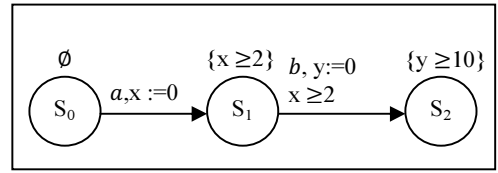

Fig .1 DATA*

The durations associated to the actions are represented by constraints on the transitions and in targets states of each one. In this sense, any enabled transition represents the beginning of the action execution. On the target state of transition, a timed expression means that the action is possibly under execution. From operational point of view, each clock is associated to an action. This clock is reset to 0 at the start of the action and will be used in the construction of the temporal constraints as guard of the transitions.

Fig.1 presents a system of two consecutives actions $a$ and $b$, the clock x is associated to the action $a$, on the locality $s_1$ the temporal expression $\{x \geq 2\}$ represents the duration of action $a$. The end of the execution of an action is deduced implicitly in the case of an action that it is causally dependent.

The action $b$ depends on $a$, so the transition is guarded by the relative duration constraint of $a$.

**Formalization:**

*Definition 1 :* a DATA* $A$ is a tuple $(L, L_0, X, T_D, L_S)$ over $ACT$ a finite set of actions, $L$ is a finite set of states, $l_0 \in L$ is the initial state, $X$ is a finite set of variables named clocks and $T_D$ is a set of edges. A subset of $L$ noted $L_f$ for terminal states (final states).

An edge $e= (l, G, a, x, l')$ represents a transition from location $l$ to location $l'$ on input symbol $a$, $x$ is a clock which is going to be reset with this transition. *G* is the corresponding guard which must be satisfied to launch this transition.

Finally, $L_S : L \to 2^{C(X)}_{fn}$ is a maximality function which decorates each state by a set of timed formula named actions durations; these actions are potentially in execution on it.

*Definition 2*: The semantic of a DATA* *A* is defined by associating to it a timed transitions system $S_A$ over $ACT \cup R^+$. A state of $S_A$ (or configuration) is a pair <*l,v*> such as *l* is a state of *A* and *v* is a valuation over *X*.

A valuation *v* is a mapping on *X* to $R^+$. Let *x* be a clock, the valuation $v[x \leftarrow 0]$ resets clock *x* to 0 and each other clock *y* to *v(y)*. The valuation *v+d* maps every clock *y* to *v(y)+d* ($d \in R^+$). A configuration <$l_0,v_0$> is initial if $l_0$ is the initial state of *A* and $\forall x \in X$, $v_0(x)$=0.

Two types of transitions between configurations of $S_A$ are possible and correspond respectively to time passing thus the run of transition from *A*.

### 3.2 Timed Refusals Region Graph

Timed refusals region graph (*TRRG*) is generated from DATA*. It decorates aggregate regions automaton with refusals.

The aggregate regions automaton [9] provides a finite abstraction of DATA*; it consists of partitioning the states space into finite regions for reducing the combinatorial explosion of regions. A region is a symbolic representation of a clock valuations sets, regrouped in equivalence classes on clock valuations. The region concept was proposed by Alur and Dill in [1].

The proposed testing model (TRRG) introduces tow kinds of refusals on aggregate regions automaton, in addition to the classical refusals named forbidden actions (*Forb*). *Forb* is defined as a set of actions which cannot be permitted from one state. However, the two new kinds of refusals are named: permanent and temporary refusals.

The *Permanent refusals* are generated by the non-determinism in system behavior after the operation of determinization. The *temporary refusals* are provoked by actions which elapsed in time.

**Formalization:**

*Definition 3:* Let $D = (L, l_0, X, T_D, L_S)$ be a DATA* over *ACT*. Its aggregate regions automaton $ARA(D) = (S, s_0, T_R)$ over *ACT*, is defined as follows:

All states of ARA(D) are of the form $s_{ij} = (l_i, r_j)$ where $l_i$ is a state and $r_j$ is a clock region.

- The set of localities is noted S. The initial locality is $s_0 = (l_0, r_0)$.
- The set of transitions $T_R$ is,

$$T_R = \left\{ t'/t' = (l,r) \xrightarrow{a} (l',r') \middle| \begin{array}{l} \exists l \xrightarrow{g,a,x} l' \in T_D \text{ and } \exists r'' \in succ(r) \\ \text{such as } r \subseteq g \text{ and } r' = r''[x \leftarrow 0] \end{array} \right\} \quad (1)$$

$s_{ij}^f = (l_i, r_j)$ is a terminal locality iff $l_i \in L_f$, and succ(r) is the set of all successors of the region r by lapsing time.

Timed refusals region graph (TRRG) extends aggregate regions automaton by sets of refusals defined as follows:

*Definition 4:* A timed refusals region graph of a DATA* *D*, (*TRRG(D)*) is a deterministic bi-labeled graph structure *A* constructed on the aggregate regions automaton of *D*, defined as structure $(S, s_0, T_R, Ref_{TPR})$.

$$A = TRRG(D) = TRRG(ARA(D)) \quad (2)$$

With: $Ref_{TPR} : S \to P\left(P(\bar{V} \cup \bar{\bar{V}})\right)$ an application that associates for any s ∈ S a set of refusals where:

$\bar{v} = \{\bar{a}(g) : a \in ACT, g \text{ is guard which must be satisfied to execute action } a\}$ and

$\bar{\bar{v}} = \{\bar{\bar{a}}(g) \in Act, g \text{ is guard which must be satisfied to execute action } a\}$ (3)

The semantic of $P\left(P(\bar{V} \cup \bar{\bar{V}})\right)$ is as follows:

$\bar{V} = \bar{a}(g) \in Ref_{TPR}(s)$: Permanent refusals means that the action *a* may be refused permanently form the state *s*, this refusal is possible but not certain. This certitude will take place after the satisfaction of guard *g*.

$\bar{\bar{V}} = \bar{\bar{a}}(g) \in Ref_{TPR}(s)$: Temporary refusals means that actions are refused as much as the guard *g* is not satisfied.

The determinization of TRRG is done according to the principle detailed in [14]; recall here that a determinization method is inspired from the classical one named subset construction.

As illustration, let consider the DATA* M of coffee machine depicted by Fig.2

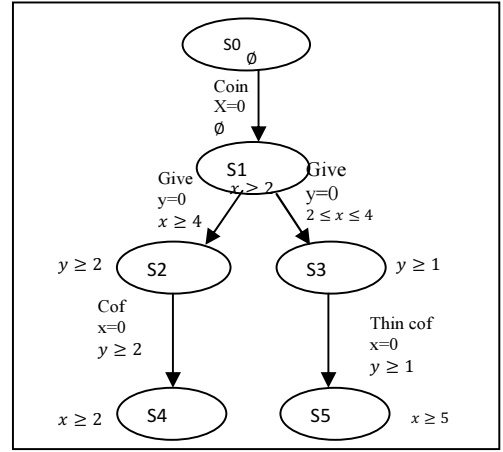

Fig .2 DATA* M of coffee machine

The TTRG M' associated to DATA* M is depicted by Fig.3

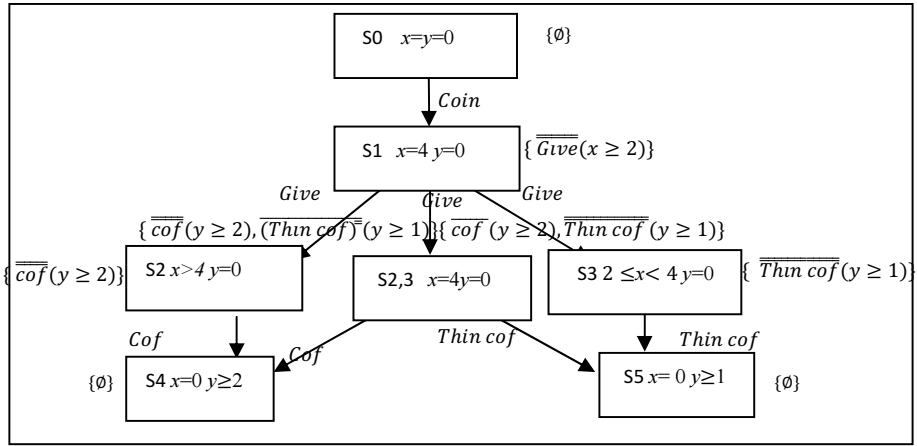

Fig .3 TRRG M' associated to DATA* M

**Timed Refusals Region Graph Generation:** Based on definitions above, a framework to create a TRRG structure is done in five (05) steps:
   Input: Specification modeled in DATA*.
   1- Determinate the DATA* structure,
   2- Compute permanent refusals for all states,
   3- Compute temporary refusals for all states,
   4- Decorate every state by sets of refusals (Forbidden, Permanent, and Temporary).
   5- Calculate the aggregate regions automaton
   Output: TRRG of the specification.

## 3.3 Canonical Tester

A canonical tester is able to detect every implementation that disagrees with a specification, thus if the implementation refuses an action after each timed trace, the specification also refuses this action. That means $I$ and $S$ have the same timed traces and the same refusals sets. This theoretical approach is necessary to generate a canonical tester. In the proposed canonical tester, three verdicts $\{pass, incon, fail\}$ are used. At every step of the test computation if a locality is reachable so it is decorated by *pass* verdict. The inconclusive verdict *incon* is produced by the non-determinism present in the system, and captured by permanent refusals set. *Fail* is a new locality introduced to canalize transitions labeled by actions which are not permitted. Two cases of actions are not allowed: first, when an action is in the forbidden set of state. The second case, when an action is offered without respecting the guard. This action is in temporary refusals set.

A framework for creating the canonical tester of DATA* specification, takes as input the TRRG and generates test cases. Localities of the test case correspond to sets of localities of the TRRG graph and edges are labeled by actions in *ACT*. Therefore, all traces of the canonical tester will be considered as test cases. The concrete timed trace can be calculated by choosing specific time points in regions.

Fig.4 presents an example of a test case associated to the canonical tester generated from TTRG M'.

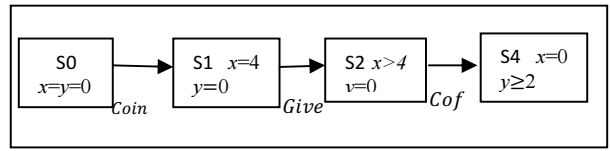

Fig .4 A Test case

## 3.4 Graph Transformation

A transformation between models is the automatic generation of a target model from a source model. This task requires a set of rules that describe how one or more constructs in the source language can be transformed to one or more constructs in the target language.

Graph Grammars [11] are used for model transformation. They are composed of production rules; each one have graphs in their left and right hand sides (LHS and RHS) (Fig.5). Rules are compared with an input graph called host graph. If a matching is found between the LHS of a rule and a sub graph in the host graph, then the rule can be applied and the matching sub graph of the host graph is replaced by the RHS of the rule. Furthermore, rules may also have a condition that must be satisfied in order to apply the rule, as well as actions to be performed when the rule is executed. A rewriting system iteratively applies matching rules in the grammar to the host graph, until no more rules are applicable. AToM[3] [2] is a graph transformation tool among others, it is implemented in the language Python. In this paper we use it.

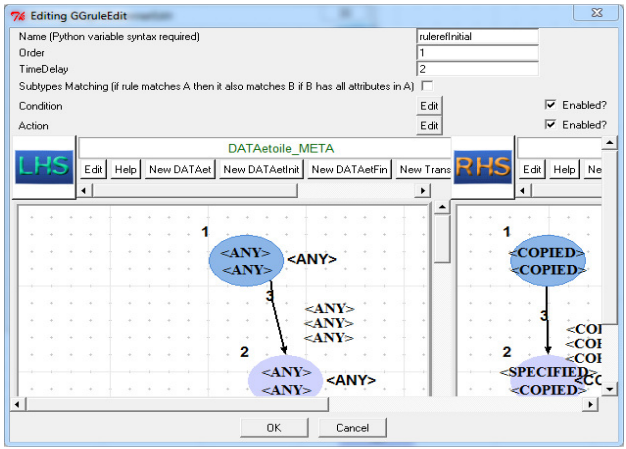

Fig. 5 A grammar rule (LHS and RHS)

## 4. The Approach

In order to test DATA* structures with a high number of states, we propose firstly a program written in python language that transforms a DATA* structure, presented as a dotty file, to a DATA* structure written in a python file which respect the syntax of AToM[3]. Also we define two meta-models; the first one associated to the DATA* model and the second one is associated with both TRRG and the canonical tester structures. We note here that the meta-models are described using UML class diagrams. Then we propose a grammar which transform DATA* to the canonical tester using TRRG for generating test cases. Meta-models and grammar are implemented in AToM[3] using python language.

### 4.1 Generation of a DATA* Respecting the Syntax of AToM[3]

Fig.6 presents an example of DATA* structure with the graph editor dotty, the translation from a dotty representation to a python representation (Figure 7.b) is done by the python program 'D_Dotty2D_Python.py' (Figure 7.a).
A graphical representation of DATA* A with AToM[3] is presented in Fig. 8.

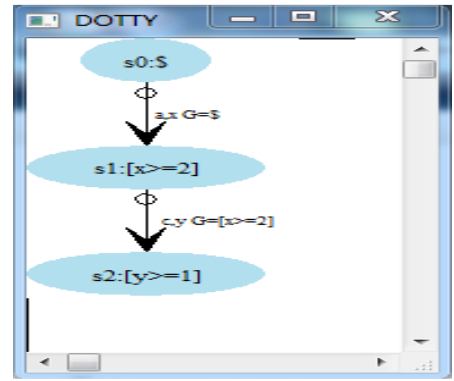

Fig 6 A dotty representation of a DATA* A

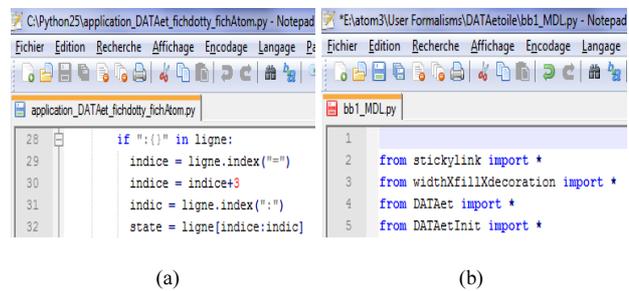

(a)          (b)

Fig. 7 Translation step (dotty-python) of a DATA* A

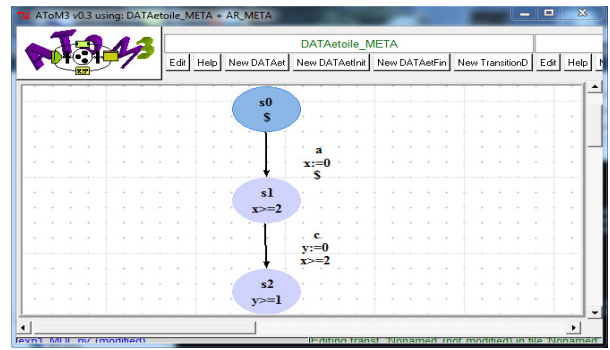

Fig 8 A graphical representation of a DATA* A with AToM[3]

### 4.2 DATA* Meta-Model

The first meta-model proposed is a class diagram composed of the following classes (Fig.9):
- *DATAet class*: represents the states of DATA*, each state has three attributes: a name (name), duration conditions (CD) and set of refusals (refusal).
- *TransitionD association*: represents the transitions of DATA*, each transition is identified by an action, a clock and a guard.
- *DATAetInit class*: represents the initial state of DATA*, it inherits attributes from *DATAet class*.

- *DATAetFin class*: represents the final state of DATA*, it inherits attributes from DATAet class.

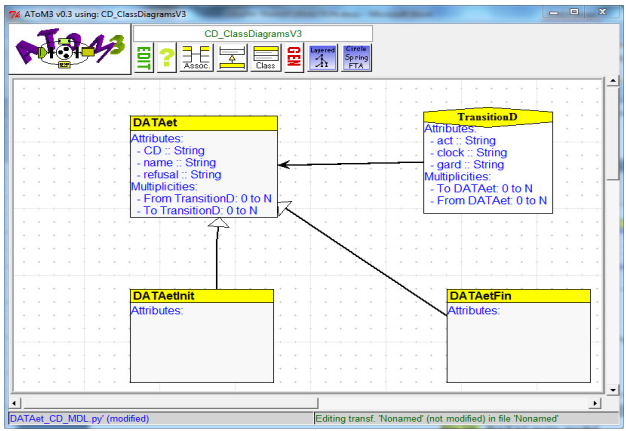

Fig. 9 DATA* meta-model

### 4.3 TRRG and Canonical Tester Meta-Model

The second meta-model describes the TRRG and the canonical tester structures. In practical point of view, they have the same structure even if they differ semantically. This meta-model is a class diagram composed of the following classes (Fig. 10):

- *TRRG_Canonical_state class*: represents the localities of TRRG and canonical tester structures, each locality has three attributes: a name (name), a clock region (clock_region) and set of refusals (refusal).

- *TRRG_Canonical_transition association*: represents the transitions of TRRG and canonical tester, each transition is identified by an action.

- *TRRG_Canonical_StateInit class*: represents the initial locality of TRRG and canonical tester; it inherits attributes from *TRRG_Canonical_state class*.

- *TRRG_Canonical_StateFin class*: represents the final locality of TRRG and canonical tester; it inherits attributes from *TRRG_Canonical_state class*.

Each class has an only graphical appearance.

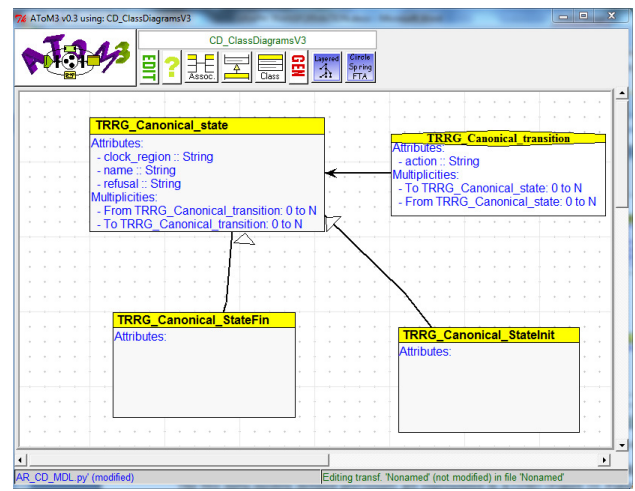

Fig. 10 TRRG and Canonical Tester Meta-Model

### 4.4 Modeling Tool (Data*, TRRG and the Canonical Tester)

The two meta-models defined previously are created in AToM$^3$ (Fig. 9, Fig. 10). They allow the generation of tool for modeling systems in DATA*, TRRG and in the canonical tester (Fig. 11).

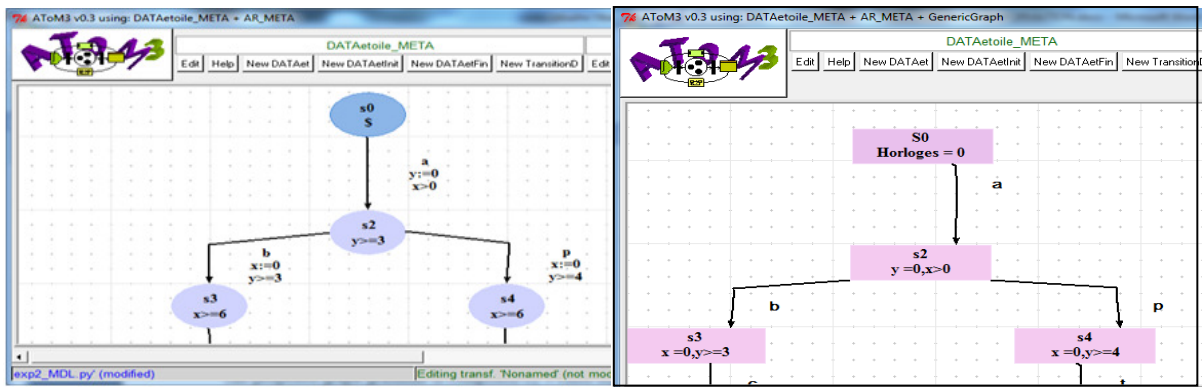

Fig. 11 Modeling tool of DATA* TRRG and Canonical Tester

### 4.4 Graph Grammar

The proposed graph grammar is composed by 23 rules organized in 2 categories (Fig. 12).

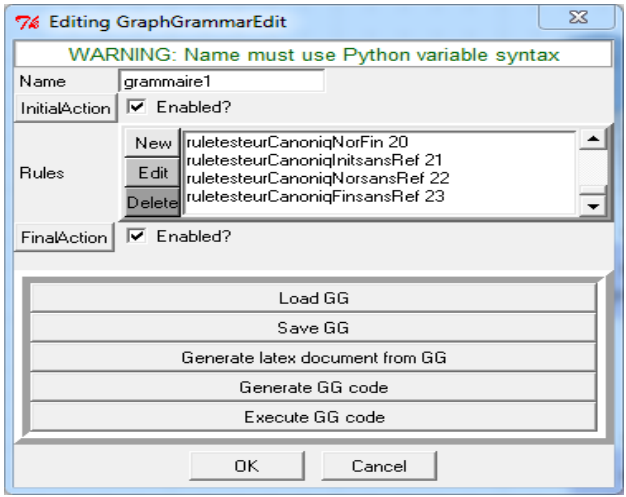

Fig. 12 Graph Grammar

The first category contains rules from 1 to 16; these rules allow the construction of TRRG based on the principle detailed in section 3.2.
- The 1st rule is used to calculate the set of refusals associated to the initial state of DATA*.
- Rules 2, determinizes and calculates the refusals set in the case of non-deterministic system.
- Rules 3 and 4 calculate respectively refusals of the rest and a final state of DATA*.
- Rule 5 is used to generate the first locality of TRRG associated to the initial state of DATA* where all clocks are reset to zero.
- Rules 6 and 7 generate the rest of TRRG.
- Rules 8 and 9 generate localities of TRRG in case of non-deterministic system.
- Rule 10 generates the final locality of TRRG associated to the final state of DATA*.
- Rules 11, 12 and 13 eliminate a generic links between DATA* and TRRG.
- Rules 14, 15 and 16 eliminate the graphical representation of DATA* model.
- The second category contains rules from 17 to 23; these rules allow the construction of canonical tester and the generation of test cases based on the principle detailed in section 3.3.

- Rule 17 is used to generate the first locality of canonical tester associated to the initial locality of TRRG.
- Rule 18 generates localities of canonical tester in case of non-deterministic system.
- Rule 19 generates the rest of canonical tester.
- Rule 20 generates the final locality of canonical tester associated to the final locality of TRRG (Fig.13).

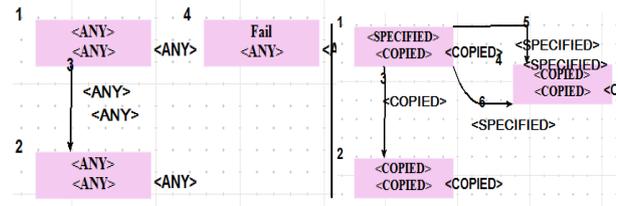

Fig. 13 Generating the final locality of canonical tester (Rule 20)

- Rules 21, 22 and 23 eliminate refusals of all localities and generate test cases.

## 5. EXAMPLE

To illustrate our approach we propose the example of the ticket reservation system "TRS". This example supposes that to buy a ticket, we generally pass by two counters. The first counter R is for making a reservation and the second counter C is for paying and taking the ticket. This agency has one waiting room, three counters of type R and two of type C.

On arrival, the client goes to the waiting room, when a counter of type R is free, he can make a reservation. Once the operation is complete, he waits until a counter C becomes free for paying and taking the ticket.

Fig. 14 presents a DATA* of TRS for two clients with the graph editor dotty.

The mapping of this DATA* with the graph editor dotty to the equivalent DATA* model of Fig. 15 is performed using python program.

We have applied our tool on the DATA* model and obtained automatically the TRRG (Fig.16), the canonical tester (Fig.17) and we have selected an example of test case (Fig.18).

Fig. 14 DATA* of TRS with the graph editor dotty

Fig. 15 DATA* of TRS with AToM$^3$

Fig. 16 TRRG associated to the DATA* of TRS

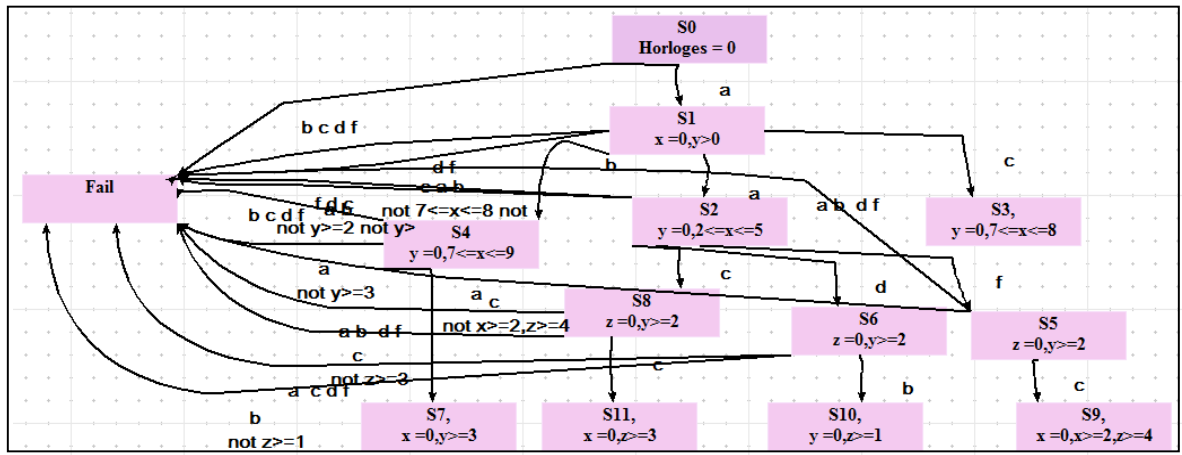

Fig. 17 The canonical tester

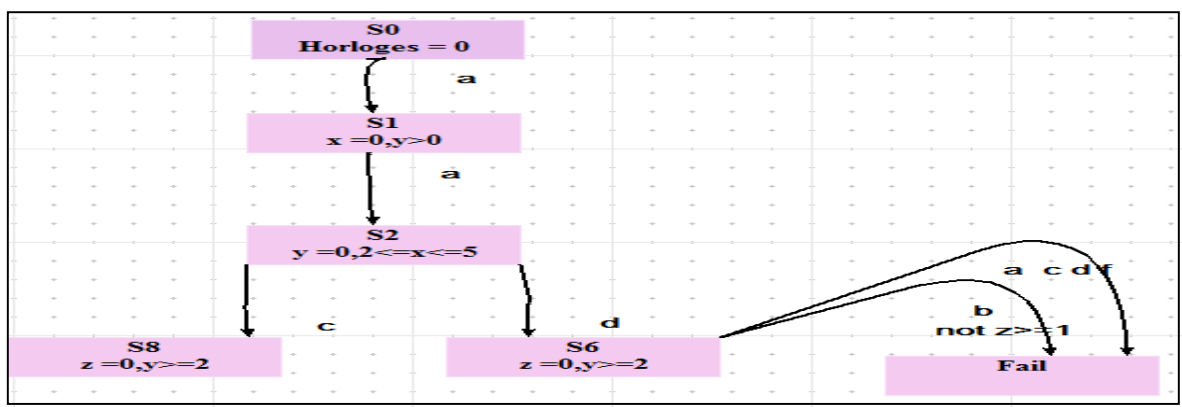

Fig. 18 Example of test case

## 6. Conclusion

In this paper, we have proposed an approach for testing timed systems, firstly, we have proposed a program written in python language that transforms a DATA* structure with a high number of states, presented as a dotty file, to a DATA* structure written in the form of a python file respecting the syntax of AToM$^3$. Secondly we have generated automatically a visual modeling tool for DATA*, TRRG and the canonical tester. The cost of building a visual modeling tool from scratch is prohibitive. Meta-modeling approach is useful to deal with this problem since it allows the modeling of the formalisms themselves. By means of graph grammars, models manipulations are expressed on a formal basis and in a graphical way. In our approach, the UML class diagram formalism is used as meta-formalism to propose a meta-model of DATA*, TRRG and the canonical tester. The meta-modeling tool AToM$^3$ is used to generate a visual modeling tool according to the proposed meta-models.

We have also proposed a graph grammar to transform a DATA* into a TRRG and into a canonical tester in order to generate test cases.

As perspectives, we plan to complete this work by strategy for choosing which of test cases are sufficient for insuring some completeness guarantees. A related problem is how to measure the "goodness" of a set of test cases and how to select test suites with some good coverage measure. We plan also to implement our approach with other tools as AGG in order to compare performances

**Hiba Hachichi** received her master's degree in computing sciences from University of Mentouri Constantine, Algeria in 2009. Currently, she is a PhD student at CFSC research group of MISC laboratory, Mentouri University of Constantine, Algeria. Her research interests are graph transformation and formal methods for verifying and testing real time systems.

**Ilham Kitouni** obtained her BEng degree from University of Mentouri Constantine, Algeria, in 1992, after 15 years in different Algerian company as head of department of Computer Sciences, she recovers CFSC research group of MISC laboratory, Mentouri University of Constantine, Algeria. From October 2009, she prepares a PhD thesis. Her research domain is formal models for real-time systems specification and validation.

**Kenza Bouaroudj** received her master's degree in computing sciences from University of Mentouri Constantine, Algeria in 2010. Currently, she is a PhD student at CFSC research group of MISC laboratory, Mentouri University of Constantine, Algeria. Her research interests are system validation and testing real-time stochastic systems.

**Djamel-eddine Saidouni** received his PHD degree in theoretical computer science from the university Paul Sabatier of Toulouse, France in 1996. Actually he is a professor at the department of computer science, Mentouri University of Constantine,
Algeria. Also, he is the head of the CFSC research group of MISC laboratory. His main research domain interests formal models for specifying and verifying critical systems, real time systems, true concurrency models and state space explosion problem.